\documentclass[aps,showpacs,twocolumn]{revtex4}
\usepackage{amsfonts}
\usepackage{color}
\begin{document}
\def\be{\begin{equation}}
\def\ee{\end{equation}}


\title{Perturbed Nonlinear Models from Noncommutativity}
\author{I. Cabrera-Carnero}\email{cabrera@cpd.ufmt.br}
\author{Luis Alejandro Correa-Borbonet}\email{borbonet@cpd.ufmt.br}
\author{G. C. S. Valadares}\email{georgevaladares@cpd.ufmt.br}
\affiliation{ Departamento de F\'{\i}sica-ICET \\
Universidade Federal de Mato Grosso\\
Av. Fernando Corr\^{e}a da Costa, $s/n^o$-Bairro Coxip\'{o}\\
78060-900-Cuiab\'{a}-MT-Brasil}
\begin{abstract}
 By means of the Eh\-ren\-fest's Theorem inside the context of a
noncommutative Quantum Mechanics it is obtained the Newton's Second
Law in noncommutative space. Considering discrete systems with
infinite degrees of freedom whose dynamical evolutions are governed
by the noncommutative Newton's Second Law we have constructed some
alternative noncommutative generalizations of two-dimensional field
theories.
\end{abstract}

\maketitle

\section{Introduction}

In the last years the research work on noncommutative (nc)
space-times have been quite extensive. Particularly, the
investigations in  Noncommutative Field Theories (NCFT) have been
very active since the appearance of these theo\-ries as low-energy
limits of string theories in the pre\-sen\-ce of magnetic fields
\cite{stringnc}.

 Besides the theoretical research in NCFT, the formulation of Quantum Mechanics on these
nc spaces(NCQM) have been considered by different authors
\cite{chaichian, poly, horvathy,hu}. It is worthy to mention that in most of
the cases treated it is considered that the time coordinate commutes
with the space coordinates. At the Quantum Field Theory level this
is the usual procedure also since the theories constructed assuming
a nc time coordinate could suffer of unitary and causality problems
\cite{gomis}. On the nc two-dimensional field theories scenario the
introduction of a nc time-coordinate is compulsory. In these cases
the nc extensions are usually achieved introducing the Moyal product
on the original action or as in the case of integrable models
starting from a ge\-ne\-ra\-li\-za\-tion of the zero curvature
condition \cite{marcoyo} \footnote{See also \cite{yo1, yo2} and
references there in.}. In order to avoid the non-unitarity in the
two-dimensional case, Euclidean models can be considered.

Obviously the existence of a noncom\-mu\-ta\-ti\-ve space-time at
quantum level could have consequences at the classical level too.
The classical limit of NCQM with commuting-time has been
investigated so far by different authors \cite{hu,acatrinei,
romero,mirza}. In \cite{acatrinei, romero, mirza} the classical
limit is taken at the level of the commutators what lead to
"deformed" Poisson brackets. Thus, the time evolution of the
coordinates is governed by a deformed Newtonian mechanics. On the
other hand in \cite{hu} the evolution of the expectation value of
the coordinates was investigated using the Ehrenfest's theorem and
the classical limit was shown to be non-Newtonian as well.
Nevertheless, in that paper, the time evolution of the momenta was
not analyzed. Additionally,  using this deformed Newtonian
Mechanics, some single particle traditional models with a finite
number of degrees of freedom have been studied, such as the
three-dimensional harmonic oscillator and the Kepler problem
\cite{romero,mirza,romero2}.

Hence, in this letter we will initially investigate in more detail
the Ehrenfest's Theorem inside the context of the NCQM with
commuting-time for the coordinates as well as for the momenta.
Secondly, taking as a point of departure this deformed Newtonian
Mechanics, we will study some mechanical models related to
two-dimensional field theories. For this purpose we will follow the
traditional approach. That is to approximate the continuous system
by a corresponding one containing discrete particles that obey the
deformed Newton's Second Law and then apply the continuous limit. In
this way it is possible to obtain alternative noncommutative
extensions of known two-dimensional field theories where time
commutes.

\section{The Newton's Second Law on noncommu\-ta\-ti\-ve spa\-ce from the Eh\-renfest's Theo\-rem}

An alternative way of investigating the classical limit of a
quantum theory is provided by the Ehrenfest's Theorem which states
that quantum expectation values of operators follow a classical
equation of motion. As we will see the classical equation of
motion followed by the expectation values of the operators
corresponding to NCQM is not Newtonian and the correction term
coincides with the one obtained in \cite{acatrinei, romero,mirza}.

Let us consider a system with $n$  degrees of freedom inside the
context of Noncommutative Quantum Mechanics with time a commuting
coordinate. This system it is describable for canonical coordinates
$(\hat{x}_i, \hat{p}_i)$ which obey the commutation relations,
\begin{eqnarray}
\left[\hat{x}_i,\hat{x}_j\right]=i\theta_{ij}, \quad
\left[\hat{x}_i,\hat{p}_j\right]= i\hbar \delta_{ij}, \quad
\left[\hat{p}_i,\hat{p}_j \right]=0,
\end{eqnarray}
with $\theta_{ij}$ an antisymmetric tensor with dimensions of
$length^2$. We will consider such systems where the Hilbert space
is the same of the corresponding commutative system
\cite{chaichian}. In this sense the dynamical evolution of the
state $|\Psi>$ is given by the usual Schr\"{o}dinger equation \be
i\hbar \frac{\partial}{\partial t} |\Psi>=H |\Psi>, \ee but the
Hamiltonian will depend on the nc coordinates, \be H=\frac{\hat{p}
\cdot \hat{p}}{2m} + V(\hat{x}).\ee
 It is possible to define a new coordinate system
 \begin{eqnarray}
\tilde{x}_i=\hat{x}_i+\frac{1}{2\hbar}\theta_{ij} \hat{p}_j, \quad
\mathrm{and} \quad \tilde{p}_i=\hat{p}_i, \label{vary}
 \end{eqnarray}
where the new coordinates satisfy the usual canonical commutation
relations,
\begin{eqnarray}
\left[\tilde{x}_i,\tilde{x}_j\right]=0, \quad
\left[\tilde{x}_i,\tilde{p}_j\right]= i\hbar \delta_{ij}, \quad
\left[\tilde{p}_i,\tilde{p}_j \right]=0.
\end{eqnarray}
The Hamiltonian in these new coordinates will read \be
H=\frac{\tilde{p}_i^2 }{2m} +
V(\tilde{x}_i-\frac{1}{2\hbar}\theta_{ij} \tilde{p}_j).
\label{hamil} \ee If we consider that the noncommutative parameter
is small, the potential can be expanded as \be V=V(\tilde{x}_i)-
\frac{1}{2\hbar}\theta_{ij} \tilde{p}_j \frac{\partial V}{\partial
\tilde{x}_i}+O (\theta^2). \label{potential} \ee Now we would like
to investigate how the Ehrenfest's theorem of Quantum Mechanics
would look in this context. For this reason let us consider that we
can still use the usual definition of mean value, because we can
consider that the noncommutativity enters as a perturbation. Since
the mean value of the time derivative of any operator ${\cal O}$
satisfies the relation,
\begin{eqnarray}
<\frac{d{\cal O }}{dt}>=\frac{i}{\hbar}<[H,{\cal O}]> ,
 \end{eqnarray}
it is obtained,
 \begin{eqnarray}
\frac{d}{dt} <\tilde{x}_i>&=&\frac{<\tilde{p}_i>}{m}-\frac{\theta_{ji}}{2\hbar}<\frac{\partial V}{\partial \tilde{x}_j}>,  \label{ehrenfest}\\
\frac{d}{dt} <\tilde{p}_i>&=& <-\frac{\partial V}{\partial
\tilde{x}_i}>+\frac{\theta_{kj}}{2\hbar}<\tilde{p}_j\frac{\partial^2
V}{\partial \tilde{x}_k \partial \tilde{x}_i}>, \nonumber
 \end{eqnarray}
which are the relations corresponding to the Ehrenfest's theorem up
to first order in $\theta$ and they show us that the mean values of
the coordinates and momentum should follow classical equations inside the
context of a deformed Newtonian mechanics. In fact the classical
limit will exists only if the noncommutative parameter $\theta_{ij}$
goes to zero at least as fast as $\hbar \rightarrow 0$, and only for
$\theta_{ij}/\hbar\rightarrow 0$ as $\theta_{ij}\rightarrow 0$ this
limit will lead to the usual Newtonian Mechanics. In \cite{romero} were derived the classical
equations
\begin{eqnarray}
\frac{d}{dt} x_i=\frac{p_i}{m}-\tilde{\theta}_{ji}\frac{\partial
V}{\partial x_j},  \quad \mathrm{and} \quad \frac{d}{dt} p_i=
-\frac{\partial V}{\partial x_i}, \label{hamiltoneqromero}
 \end{eqnarray}
 with the coordinates satisfying the Poisson brackets,
\begin{eqnarray}
\{x_i, x_j\}=i\tilde{\theta}_{ij}, \quad \{x_i, p_j\}=
\delta_{ij}, \quad \{p_i,p_j \}=0. \label{poisson}
\end{eqnarray}
After introducing the variables
$\tilde{x}_i=x_i+\frac{1}{2}\tilde{\theta}_{ij} p_j, \quad
\tilde{p}_i=p_i$ the equations (\ref{hamiltoneqromero}) can be
written as
 \begin{eqnarray}
\frac{d}{dt}
\tilde{x}_i=&&\frac{\tilde{p}_i}{m}-\frac{\tilde{\theta}_{ji}}{2}\frac{\partial
V}{\partial \tilde{x}_j},  \nonumber \\ \frac{d}{dt}
\tilde{p}_i=&& -\frac{\partial V}{\partial
\tilde{x}_i}+\frac{\tilde{\theta}_{kj}}{2}\tilde{p}_j\frac{\partial^2
V}{\partial \tilde{x}_k \partial \tilde{x}_i},
\label{hamiltontilvariables}
 \end{eqnarray}
up to first order in $\tilde{\theta}_{ij}$.
The equivalence of (\ref{ehrenfest}) and
(\ref{hamiltontilvariables}) can be proven if we consider that the
classical limit of (\ref{ehrenfest}) exists since, in principle, we
can take $\theta_{ij}=\hbar\tilde{\theta}_{ij}$ in (\ref{ehrenfest}).
Moreover if we assume that the wave function of our system is
different from zero only in a small region of space around the
mean value of the coordinate, say $<{\bf x}>$ we can expand the
derivative of the potential around the mean values
$\overline{\tilde{x}}_i=<\tilde{x}_i>$. In this way we end up with
relations analog to (\ref{hamiltontilvariables}) for the mean
values, i.e.,
\begin{eqnarray}
\frac{d}{dt} \overline{\tilde{x}}_i=&&\frac{\overline
{\tilde{p}}_i}{m}- \frac{\tilde{\theta}_{ji}}{2}\frac{\partial
V}{\partial \overline{\tilde{x}}_j} \nonumber \\ \frac{d}{dt}
\overline{\tilde{p}}_i=&& -\frac{\partial V}{\partial
\overline{\tilde{x}}_i}+\frac{\tilde{\theta}_{kj}}{2}\overline{\tilde{p}}_j\frac{\partial^2
V}{\partial \overline{\tilde{x}}_k \partial
\overline{\tilde{x}}_i}, \label{ehrenfest2}
\end{eqnarray}
which are valid as far as  $|\frac{\partial V}{\partial
\overline{\tilde{x}}_i}|\gg \frac{1}{2}|\frac{\partial^3
V}{\partial \overline{\tilde{x}}_k \overline{\tilde{x}}_j
\overline{\tilde{x}}_i}|<\Delta \tilde{x}_j \Delta \tilde{x}_k>$.
These inequalities are better satisfied when the potential changes
smoothly with the coordinates $\tilde{x}_i$ and when the spatial
extension of the wave packet is small. But as small values of
$<\Delta \tilde{x}>$ lead to a large indeterminacy in the value of
the momentum in order that one may approximately apply classical
ideas about the motion of a particle
 it is necessary also to consider that $ <\frac{\tilde{p}_i^2}{2\mu}>\gg \frac{<(\Delta
\tilde{p}_i)^2>}{2 \mu}$ holds. Thus these conditions are better
satisfied when the particle is moving with a large momentum in a
smoothly varying external field. And in a situation like this the
center of the wave packet could move classically following a
deformed Newton's Second law,
\begin{eqnarray}
m \frac{d^2}{dt^2}\overline{\hat{x}}_i=-\frac{\partial V}{\partial
\overline{\hat{x}}_i}+m \tilde{\theta}_{ij}\frac{\partial^2
V}{\partial \overline{\hat{x}}_k
\partial \overline{\hat{x}}_j} \frac{d}{dt}{\overline{\hat{x}}}_k, \label{Newton}
\end{eqnarray}
which is obtained from (\ref{ehrenfest2}) after returning to the
original variables (\ref{vary}) and which is the Newton's Second
Law in nc space obtained in \cite{acatrinei, romero, mirza}, expressed now for the mean values. Note
that for a single particle constrained to move in only one spatial
dimension the nc term will not be present. So in principle in
order to consider nc two-dimensional models one will be forced to
introduce the noncommutativity of the time coordinate.

Using this deformed Newton's Second Law (\ref{Newton}) dif\-fe\-rent
traditional single particle problems have been studied like the
central forced potential \cite{mirza}, the three-dimensional
harmonic oscillator \cite{romero} and the Kepler pro\-blem
\cite{romero2}. In the following we would like to investigate the
possible formulation of some discrete models in this setup and
their corresponding continuous limits such that alternative
nc extensions of two-dimensional field theories will be obtained.

\section{Infinite Os\-ci\-lla\-tors Chains and the Newton's Second Law on NC Space}

The equation (\ref{Newton}), as it was derived applying the
Ehrenfest's theorem, will describe the dynamical evolution of a
single particle with $n$ degrees of freedom
whose spatial coordinates do not commute. Our concern now is with a
discrete system with $n \rightarrow \infty$ degrees of freedom but
now formed by an infinite amount of interacting particles in such a
way that it approximates an infinitely long rod that can undergo
small longitudinal vibrations.  So this system it is composed of a
one-di\-men\-sio\-nal infinite chain of equal mass points
interacting particles spaced a distance $\delta$ apart. But now, let
us consider that the coordinates of the point particles do not
commute at quantum level. In this sense we have extended the
concept of noncommutativity from coordinates to degrees of
freedom. Consequently if we are interested on the classical
description of this system we will be dealing with Poisson
brackets like (\ref{poisson}). And in this sense the system will evolve in time following the
 Newton's Second law on nc space,
\begin{eqnarray}
m\frac{d^2}{dt^2}x_i=-\frac{\partial V}{\partial x_i}+m
\tilde{\theta}_{ij} \frac{\partial^2 V}{\partial x_k
\partial x_j}\frac{d}{dt} x_k. \label{Newtonn}
\end{eqnarray}
This consideration can be related to the fact that for example for
two-dimensional field theories it is valid that that the
Moyal bracket of a function and the same function
spatial-translated is different from zero, i.e. $
\left[\phi(x,t),\phi(x+a,t)\right]_{\star}\neq 0 $. The
description of the continuous system will be obtained as the
continuous limit of the discrete system is approached. So
initially let us consider elastic interactions and in this sense
the corres\-ponding Hamiltonian will be
\begin{equation}
H=\frac{1}{2}\sum_{i} \Big[ m \dot{x}_i^2 + k(x_{i+1}-x_i)^2 \Big],
\end{equation}
where we have assumed that the mass points can move only along the
length of the chain and we have denoted the displacement of the
$i$th particle by $x_i$. Here the $\cdot$ corresponds to
time-derivative. We are going to consider that only the
coordinates of the nearest neighbors do not commute, so
\begin{eqnarray}
\tilde{\theta}_{ij}=\left( \begin{array}{cc}
               \theta & \mathrm{for} \quad j=i+1, \\
               -\theta & \mathrm{for} \quad j=i-1, \\
                 0    & \mathrm{otherwise}. \quad \quad
              \end{array} \right.,
\end{eqnarray}
with $\theta$ a constant parameter.
Hence the dynamical evolution of these coordinates go\-ver\-ned by
(\ref{Newtonn}) will be
\begin{eqnarray}
m\ddot{x}_i&-&k(x_{i+1}-2x_i+x_{i-1}) \label{harmonic}\\
&-&m\theta
k(2\dot{x}_{i+1}-2\dot{x}_{i-1}-\dot{x}_{i+2}+\dot{x}_{i-2})=0.
\nonumber
\end{eqnarray}
 So even when we consider a potential with nearest
neighbors interaction and that only the nearest neighbor coordinates
don't commute we have the influen\-ce of second neighbors on the
equations of motion in the form of a dissipative term.

We can find the corresponding field theory computing the
continuum limit.  For this purpose consider that $\delta$
approaches zero and in this case the integer index $i$ identifying
the particular mass becomes the continuous position coordinate $x$
so instead of $x_i(t)$ we will have $\phi(x,t)$. Using the Taylor
expansions for
\begin{eqnarray}
x_{i\pm 1}(t) &\rightarrow& \phi(x \pm \delta,t)=\phi(x,t)\pm
\delta
\partial_x \phi(x,t)\nonumber \\
 && \hspace{2cm} +\dots + O(\delta^4) \nonumber \\
x_{i\pm 2}(t) & \rightarrow & \phi(x \pm 2 \delta,t)=\phi(x,t)\pm
2 \delta
\partial_x \phi(x,t)\nonumber \\
 && \hspace{2cm} +\dots + O(\delta^4), \label{taylor}
\end{eqnarray}
the equations of motion (\ref{harmonic}) become \be
 \frac{\partial^2\phi}{\partial t^2}-\frac{\partial^2 \phi}{\partial
 x^2}+2
 \theta \sqrt{\mu Y}\delta^2 \frac{\partial^4 \phi}{\partial x^3 \partial t}
 =0,\label{wavedeformed}
 \ee
after redefining the variables as $t \rightarrow
 t\sqrt{\frac{Y}{\mu}}$, with $\mu=\frac{m}{\delta}$ the mass per unit length and $Y=k \delta$ the Young modulus.
So the wave equation has gained a correction term. In this way it contrasts with the traditional
noncommutative extension of the massless free field done through
the introduction of the Moyal product on the action, which due to
the properties of this product the wave equation preserves its
original commutative shape \cite{micu}.

Considering the nc parameter as a small perturbation we can look
for the solution as $\phi=\phi_0+\theta \phi_1$ with $\phi_0$ a
solution of the wave equation that in general it can be expressed
as $\phi_0=f_1(x+t)+f_2(x-t)$. Introducing this expression in the
equation (\ref{wavedeformed}) we are left with \be
\partial_{tt}^2 \phi_1 - \partial_{xx}^2 \phi_1+2\theta \delta^2
\sqrt{\mu Y}
\partial^4_{xxxx}(f_1-f_2)=0, \label{phi1}\ee
at first order of $\theta$. Defining \be F(x,t)=2\theta \delta^2
\sqrt{\mu Y}
\partial^4_{xxxx}(f_1-f_2), \ee the solution of (\ref{phi1}) dependent on the initial
conditions $\phi(x,0)=u(x)$, $\phi_t(x,0)=v(x)$ can be
particularly expressed as \begin{eqnarray}
\phi_1(x,t)=&&\frac{u(x-t)+u(x+t)}{2}+\frac{1}{2}\int_{x-t}^{x+t}v(\xi)d\xi
\nonumber \\ &&+ \frac{1}{2}\int_0^t d\tau
\int_{x-(t-\tau)}^{x+(t-\tau)}F(\xi,\tau)d\xi, \end{eqnarray} for
$-\infty<x<\infty$, $0<t<\infty$ \cite{tijonov}.

Imagine now that the infinitely long elastic rod is not elastic so
we could model it as a nonlinear oscillator chain, as for example
the Fermi-Pasta-Ulam model \cite{fermi} where we can consider
additionally quadratic forces,
\be V=\sum_i \Big[
\frac{k}{2}(x_{i+1}-x_i)^2+\frac{\alpha}{3}(x_{i+1}-x_i)^3 \Big],
\ee
and it is called $\alpha$-model or additional cubic forces,
\be
V=\sum_i
\Big[\frac{k}{2}(x_{i+1}-x_i)^2+\frac{\beta}{4}(x_{i+1}-x_i)^4
\Big], \ee
and it is called $\beta$-model. These models are closely related
to many physical problems like the integrability of nonlinear
equations, their soliton solutions, dynamical chaos and
Bose-Einstein condensates. So, following the above procedure the
corresponding equations motion of these models inside the context
of a non-Newtonian Mechanics will read
\begin{eqnarray}
m\ddot{x}_i=&&k(x_{i+1}-2x_i+x_{i-1}) \quad \quad
\alpha-\mathrm{model} \nonumber
\\
&&+\alpha[(x_{i+1}-x_i)^2 -(x_{i}-x_{i-1})^2] \nonumber \\
&&+m\theta \frac{d}{dt}\{k(2x_{i+1}-2x_{i-1}+x_{i-2}-x_{i+2})  \nonumber \\
&&+\alpha[
(x_{i+1}-x_i)^2-(x_{i-1}-x_{i-2})^2 \nonumber \\ && -(x_{i+2}-x_{i+1})^2+(x_{i}-x_{i-1})^2 ] \},  \\
m\ddot{x}_i=&&k(x_{i+1}-2x_i+x_{i-1}) \quad \quad \beta-\mathrm{model} \nonumber \\
&&+\beta[(x_{i+1}-x_i)^3-(x_{i}-x_{i-1})^3] \nonumber
\\&& +m\theta \frac{d}{dt}\{
k(2x_{i+1}-2x_{i-1}+x_{i-2}-x_{i+2}) \nonumber
\\ && + \beta[-(x_{i-1}-x_{i-2})^3
-(x_{i+2}-x_{i+1})^3 \nonumber
\\ &&+(x_{i}-x_{i-1})^3+(x_{i+1}-x_{i})^3] \}.
\end{eqnarray}
If we apply the continuum limit $\delta \rightarrow 0$,
introducing the corresponding Taylor expansions (\ref{taylor}) up
to fourth order in $\delta$ we will obtain the nonlinear field
theories defined by the following partial differential equations:
\begin{eqnarray}
 && \partial^2_{tt}\phi-[1+2\bar{\alpha}\delta\partial_x\phi]\partial^2_{xx}
\phi \quad \quad \quad \quad \alpha-\mathrm{model} \nonumber
\\ && \hspace{0.7cm} -\frac{\delta^2}{12}\partial^4_{xxxx}\phi +2\theta \sqrt{\mu
Y}\delta^2\frac{d}{dt}\Big[
\partial^3_{xxx}\phi \nonumber \\ && \hspace{2.5cm} +2\delta \bar{\alpha} \partial_x(\partial_x
\phi \partial_{xx}^2 \phi)\Big]=0,  \label{alpha}
\\ && \partial^2_{tt}\phi-[1+3\bar{\beta}\delta^2 (\partial_x
\phi)^2]\partial^2_{xx} \phi
 \quad \quad \beta-\mathrm{model} \nonumber  \\  && \hspace{0.7cm} -\frac{\delta^2}{12}\partial_{xxxx}^4 \phi+2 \theta \sqrt{\mu Y}\delta^2 \partial^4_{xxxt} \phi=0,  \label{beta}
\end{eqnarray}
where we have redefined the variables $t \rightarrow
 t\sqrt{\frac{Y}{\mu}}$, $\bar{\alpha}=\frac{\alpha}{k}$ and
 $\bar{\beta}=\frac{\beta}{k}$.
Note that for the $\beta$-model the deformation term is the same
than for the harmonic oscillators which is a consequence of the
fact that the contributions of the $\beta$-terms appear only at
fifth order of $\delta$. So the application of the deformed
Newton's Second Law (\ref{Newtonn}) to the dy\-na\-mi\-cal evolution of
the system has lead to wave equations that describe shallow
water waves in classical hydrodynamics perturbed by dissipative
terms in the case of the $\beta$-model and by dissipative and
nonlinear terms in the case of the $\alpha-$ model (\ref{alpha}).

A related integrable model to the previous nonlinear systems is
the open Toda lattice \cite{Toda}. The model consists of a
one-dimensional chain of $n$ particles with exponential nonlinear
nearest neighbor interaction \cite{Toda} in such a way that the
potential is expressed as  $ V=\sum_{i}\Big[
\frac{a}{b}e^{b(x_{i+1}-x_{i})}-a(x_{i+1}-x_{i})\Big]$. If we
consider that the dynamical behavior of this system is governed by
the deformed Newton's Second Law (\ref{Newtonn}) the equations of
motion will read:
\begin{eqnarray}
m\ddot{x}_i=&&a\big[e^{x_{i+1}-x_{i}}-e^{x_{i}-x_{i-1}}\big] + m
\theta a b \frac{d}{dt}\Big[e^{x_{i+1}-x_{i}} \label{toda}
\nonumber \\ &&
+e^{x_{i}-x_{i-1}}-e^{x_{i+2}-x_{i+1}}-e^{x_{i-1}-x_{i-2}}\Big].
\end{eqnarray}
From now on, by simplicity reasons, we will assume \-$m=a=b=1$. And
once again let us follow the previous procedure in order to
compute the conti\-nuum limit of this model. For this purpose we
introduce the Taylor expansions (\ref{taylor}) up to fourth order of $\delta$ in the previous
expression and after normalizing the time $t\rightarrow t \delta$
it is obtained,
\begin{eqnarray}
\partial^2_{tt}\phi-[1+\delta
\partial_x\phi+\frac{\delta^2}{2}(\partial_x
\phi)^2]\partial^2_{xx}\phi-\frac{\delta^2}{12}\partial^4_{xxxx}
\phi  && \\+ 2\theta \delta^2\frac{d}{dt}\Big[
\partial^3_{xxx}\phi+\delta\partial_x(\partial_x \phi
\partial_{xx}^2 \phi)\Big]=0 && \nonumber
\end{eqnarray}
As for computing the continuum limit we have expanded the
exponential it is not surprising that in this case we have
obtained the same correction term than in the $\alpha$ model
(\ref{alpha}). But there is another way to compute the continuous
limit of the Toda model, for which we rewrite the equation
(\ref{toda}) as \begin{eqnarray} \ddot x_i=
e^{x_{i+1}}+e^{x_{i-1}}-2e^{x_i}&+& \theta\frac{d}{dt} \Big[
2(e^{x_{i+1}}-e^{x_{i-1}}) \nonumber
\\ &-& e^{x_{i+2}}+e^{x_{i+2}}\Big], \end{eqnarray} where we have considered that
$x_{i+1}-x_i\rightarrow x_i$. Let us introduce $e^{x_i}=1+V_i$ in
order to consider the continuous limit where $V_i(t) \rightarrow
\phi(x,t)$ with the corresponding Taylor expansions (\ref{taylor})
for $V_{i\pm1}, V_{i\pm 2}$ but now up to fourth order in $\delta$. So we arrive at \be \partial^2_{tt}
u=
\partial^2_{xx} u + \delta
\partial_{xx}^2 (u^2)+\frac{\delta^2}{12} \partial^4_{xxxx}
u-2\theta\delta^2\partial^4_{xxxt}u, \label{boussinesq}\ee with
$\phi=e^{\delta u}-1$ and where the time coordinate has been
redefined $t \rightarrow t \delta$. In this way we have obtained
the Boussinesq equation perturbed by a nc dissipative term which is
the same perturbation that appeared on the wave equation
(\ref{wavedeformed}) and in the $\beta$-model (\ref{beta}). By the
other side, it is well known that the Boussinesq equation leads to
the KdV equation when tra\-ve\-lling waves in only one direction
are considered. We will try in the following to approximately
derive from (\ref{boussinesq})
 the corresponding nc extension of the KdV equation. So we will obtain the qualitatively
 behavior of the KdV model in this nc context. For this purpose let's do the change of variables
$\varphi=\frac{d u}{dx}$ which after integration leads to
 \begin{eqnarray}
\partial^2_{tt} \varphi=\partial^2_{xx}\varphi + 2\delta
\partial_{xx}^2 \varphi\partial_{x}\varphi&+&\frac{\delta^2}{12} \partial^4_{xxxx}
\varphi \\&-&2\theta \delta^2\partial^4_{xxxt}\varphi +const.
\nonumber
\end{eqnarray}
  Let us consider that the function $\varphi$ represents a wave
  propagating to the right,
  $\varphi(x,t)=\varphi(x-vt)$, where $v \neq 0$ is a parameter proportional to the velocity of propagation of
  the wave with the extra condition $v \neq 1$ in order not to loose terms. In this sense $\partial_x \varphi=-v \partial_t \varphi$ which we
  introduce on the previous equations obtaining
  \begin{eqnarray}
(1-\frac{1}{v^2}) \partial^2_{xx}\varphi &+& 2\delta
\partial_{xx}^2 \varphi \partial_{x}\varphi+\frac{\delta^2}{12} \partial^4_{xxxx}
\varphi\nonumber
\\ &-&2\theta\frac{\delta^2}{v}\partial^4_{xxxx}\varphi +const=0.
  \end{eqnarray}
 Considering the equation for $\partial_x \varphi=\Phi$, the time
derivative can be recovered introducing a parameter $\epsilon$, $\epsilon \neq 1$
that will give the proportion of the time
derivative and the spatial derivative on the terms where
$\partial_x=-v\partial_t$ was introduced,
 \begin{eqnarray}
(1-\frac{1}{v^2})\epsilon \partial_{x}\Phi
&+&(1-\frac{1}{v^2})(1-\epsilon)
\partial_{t}\Phi \nonumber + 2\delta \Phi \partial_{x}
\Phi \\ &+&\frac{\delta^2}{12}
\partial^3_{xxx} \Phi-2\theta\frac{\delta^4}{v}\epsilon\partial^3_{xxx}\Phi\nonumber\\&-&2\theta\frac{\delta^4}{v}(1-\epsilon)\partial^3_{xxt}\Phi
+const=0 \nonumber.
  \end{eqnarray}
  Which after the transformation $\xi=x-\frac{\epsilon}{1-\epsilon}t$, $\tau=\frac{t}{1-\epsilon}$ and
  taking $const=0$ it can be cast on the form \be
\partial_{\tau}\Phi+a \Phi
\partial_{\xi} \Phi+b\partial_{\xi \xi
\xi}^3 \Phi-\theta c\partial^3_{\xi \xi \tau} \Phi=0,
 \ee
with $a=\frac{2\delta}{1-\frac{1}{v^2}}$, $b=\frac{\delta^2
}{12(1-\frac{1}{v^2})}$ and $c=\frac{2\delta^3}{(v-\frac{1}{v})}$.
Moreover scaling $\xi \rightarrow b^{-\frac{1}{3}}\xi$, $\tau
\rightarrow -\tau$ and $\Phi \rightarrow ab^{-\frac{1}{3}\Phi}$ it
can be expressed as the KdV equation  in its more usual form, now with an additional nc term,
\be
\partial_{\tau}\Phi=\Phi
\partial_{\xi} \Phi+\partial_{\xi \xi
\xi}^3 \Phi+\theta \frac{2}{v}
\left(\frac{12^2\delta^2}{1-\frac{1}{v^2}}\right)^{\frac{1}{3}}\partial^3_{\xi
\xi \tau} \Phi.
 \ee
  Considering the noncommutative parameter small the additional term can be considered as
a weak perturbation on the KdV equation. Note that in this case
this perturbation in order to be real must also satisfy $v>1$. The effects of weak
perturbations on the solutions of the KdV equation have been
studied by different methods like the proposal based on the
inverse scattering theory \cite{kaup}, or direct perturbation
theory \cite{ablowitz}, or by Green function formalism
\cite{mann}. So employing these methods would be possible to find
the solutions and in this way to investigate the influence of the
additional nc term on the soliton solutions of the original KdV
equation. At this point could be interesting to compare with other
nc extension of the KdV equation proposed on the literature
\cite{dimakiskdv,legare}. If on the proposal presented in
\cite{dimakiskdv,legare} we expand the Moyal-$star$ product on
$\theta$ we will find that the first correction term will appear
for the second order of the nc parameter $\theta$,
 \begin{eqnarray}
u_t+u_{xxx}&-&6uu_x-\frac{\theta^2}{2}\Big[\partial^2_{xt}u
\partial_{xtt}^3u   \\ &+& \partial^3_{xxx}u \partial^2_{tt}u-4\partial^3_{xxt}u \partial^2_{xt}u
\Big]=0. \nonumber
 \end{eqnarray}
Once again this result contrasts with the result obtained in this
letter where the correction term appears at first order of
$\theta$. This fact is in some sense natural since on the nc
extensions of two-dimensional integrable field theories like KdV
equation, where the nc extensions are obtained starting from the
original commutative field theory and introducing the Moyal
product on the zero-curvature condition, the time is considered a
non-commuting coordinate. By the other side in our approach time
has always been a commuting coordinate, so it is expected the non
correspondence of both formalisms.

The proper comparison must be done between the nc two-dimensional
Moyal field theories  and the field theo\-ries obtained as the
continuous limits of corres\-pon\-ding discrete models whose dynamical
evolutions will be go\-ver\-ned by a Newton's Second Law obtained from a NCQM with a non-commuting time
coordinate. In \cite{teotonio}  was proposed a NCQM with commuting time, based on the
works of Doplicher et al \cite{doplicher} and which it seems to preserve the
unitary. Of course it would be interesting to investigate this in
the future. By the moment we present these models as alternative
nc formulations of field theories which emerge from the continuous
limits of discrete models whose dynamical evolutions are governed
by the deformed Newton's Second Law (\ref{Newtonn}).

\section*{Acknowledgments}

 One of us(LACB) would like to thank the Perimeter Ins\-titute as well as the
 University of Waterloo for the kind hospitality and where the last part of this work was done. This work has been
 supported by CNPq-FAPEMAT.


\begin{thebibliography}{99}
\bibitem{stringnc}
Alain Connes, Michael R. Douglas and Albert Schwarz, JHEP {\bf
9802} (1998)
003; \\
Michael R. Douglas and Chris Hull, JHEP {\bf 9802} (1998)
008; \\
Nathan Seiberg and Edward Witten, JHEP {\bf 9909} (1999) 032.
\bibitem{chaichian}
M. Chaichian, M. M. Sheik-Jabbari and A. Tureanu, Phys. Rev. Lett.
{\bf 86} (2001) 2716.
\bibitem{poly}
V. P. Nair and A. Polychronakos, Phys.Lett. {\bf B505} (2001) 267.
\bibitem{horvathy}
C. Duval and P. A.Horváthy, Phys. Lett {\bf B479} (2000) 284; J. Phys {\bf A34} (2001) 10097.
\bibitem{hu}
A. Eftekharzadeh and B. L. Hu, Braz. Jour. Phys. {\bf 35} no. 2A
(2005) 333.
\bibitem{gomis}
 J. Gomis and T. Mehen, Nucl. Phys. {\bf B591} (2000) 265.
\bibitem{acatrinei}
Ci\-prian Aca\-tri\-nei, Pre-print [arXiv:hep-th/0106141].
\bibitem{romero}
 Juan M. Romero, J.A. Santiago and J. David Vergara, Phys. Lett {\bf A310} (2003) 9.
 \bibitem{mirza}
B. Mirza and M. Dehghani, Commun. Theor. Phys. {\bf 42} (2004)
183.
\bibitem{marcoyo}
I. Cabrera-Carnero and  M. Moriconi, Nucl.Phys. {\bf B673} (2003)
437.
\bibitem{yo1}
I. Cabrera-Carnero, JHEP {\bf 0510} (2005) 071.
\bibitem{yo2}
I. Cabrera-Carnero, J. Phys. {\bf A39} (2006)
5979.
\bibitem{romero2}
Juan M. Romero and J. David Vergara, Mod. Phys. Lett. {\bf A18}
(2003) 1673.
\bibitem{Toda}
M. Toda, J. Phys. Soc. Jap. {\bf 22} (1967) 431.
\bibitem{dimakiskdv}
Aristophanes Dimakis and Folkert Mue\-ller-Hoi\-ssen,
[arXiv:hep-th/0007074].
\bibitem{micu}
A. Micu and M.M. Sheikh-Jabbari, JHEP {\bf 0101} (2001) 025.
\bibitem{tijonov}
A. A. Samarskii and A. N. Tikhonov, "Equations of Mathematical
Physics", Dover Publications, INC. New York 1990, Copyrigth 1963
by Pergamon Press Ltd.
\bibitem{fermi}
E. Fermi, J. Pasta and S. Ulam, Los Alamos Report LA-1940 (1955).
\bibitem{legare}
M. Legare, J. Phys. A:Math. Gen.{\bf 35} (2002) 5489; Pre-print
[arXiv: hep-th/0012077].
\bibitem{kaup}
D. J. Kaup and A. C. Newell, Proc. Roy. Soc. London Ser. A {\bf
361} (1978) 413; \\
V.I. Karpman and E. M. Maslov, Sov. Phys. JETP {\bf 46} (1977)
281, Sov. Phys. JETP {\bf 48} (1978) 252.
\bibitem{ablowitz}
Y. Kodama and M. J. Ablowitz, Stud. Appl. Math. {\bf 64} (1981)
225.
\bibitem{mann}
E. Mann, J. Math. Phys. {\bf 38} (1997) 3772.
\bibitem{teotonio}
 A. P. Balachandran, A. M. Marques, A. R. Queiroz and P.
 Teo\-to\-nio-So\-bri\-nho, JHEP {\bf 0410} (2004) 072.
\bibitem{doplicher}
S. Doplicher, K. Fredenhagen and J. E. Roberts, Commun.Math.Phys.
{\bf 172} (1995) 187.

\end{thebibliography}
\end{document}